\documentclass[twocolumn]{aastex631}

\usepackage{amsmath}
\usepackage{csquotes}
\usepackage{soul}

\newcommand{{\halpha}}{H$\alpha$}
\newcommand{{\nii}}{[NII]$\lambda6584$} 

\received{January 25, 2024}
\revised{\today}
\shorttitle{Metallicity offset between clumpy and non-clumpy SFGs}
\shortauthors{Sok et al.}
\graphicspath{{./}{figures/}}

\begin{document}

\title{An Indication of Gas Inflow in Clumpy Star-Forming Galaxies near $z\sim1$:\\ Lower Gas-Phase Metallicities in Clumpy Galaxies Compared to Non-Clumpy Galaxies}

\correspondingauthor{Visal Sok}
\email{sokvisal@yorku.ca}

\author[0000-0002-0786-7307]{Visal Sok}
\affiliation{Department of Physics and Astronomy, York University, 4700 Keele St, Toronto, ON M3J 1P3, Canada}

\author[0000-0002-9330-9108]{Adam Muzzin}
\affiliation{Department of Physics and Astronomy, York University, 4700 Keele St, Toronto, ON M3J 1P3, Canada}

\author[0000-0002-9655-1063]{Pascale Jablonka}
\affiliation{Laboratoire d'astrophysique, \'{E}cole Polytechnique F\'{e}d\'{e}rale de Lausanne (EPFL), CH-1290 Sauverny, Switzerland}

\author[0000-0002-3503-8899]{Vivian Yun Yan Tan}
\affiliation{Department of Physics and Astronomy, York University, 4700 Keele St, Toronto, ON M3J 1P3, Canada}

\author[0000-0002-7248-1566]{Z. Cemile Marsan}
\affiliation{Department of Physics and Astronomy, York University, 4700 Keele St, Toronto, ON M3J 1P3, Canada}

\author[0000-0001-9002-3502]{Danilo Marchesini}
\affiliation{Department of Physics \& Astronomy, Tufts University, 419 Boston Ave, Medford, MA 02155, USA}

\author[0000-0002-6572-7089]{Gillian Wilson}
\affiliation{Department of Physics, University of California Merced, 5200 Lake Road, Merced, CA 95343, USA }

\author[0000-0002-2250-8687]{Leo Y. Alcorn}
\affiliation{David A. Dunlap Department of Astronomy \& Astrophysics, University of Toronto, 50 Saint George Street, Toronto, ON M5S 3H4, Canada}



\begin{abstract}
    Despite the ubiquity of clumpy star-forming galaxies at high-redshift, the origin of clumps are still largely unconstrained due to the limited observations that can validate the mechanisms for clump formation. We postulate that if clumps form due to the accretion of metal-poor gas that leads to violent disk instability, clumpy galaxies should have lower gas-phase metallicities compared to non-clumpy galaxies. In this work, we obtain the near-infrared spectrum for 42 clumpy and non-clumpy star-forming galaxies of similar masses, SFRs, and colors at $z\approx0.7$ using the Gemini Near-Infrared Spectrograph (GNIRS) and infer their gas-phase metallicity from the {\nii} and {\halpha} line ratio. We find that clumpy galaxies have lower metallicities compared to non-clumpy galaxies, with an offset in the weighted average metallicity of $0.07\pm0.02$ dex. We also find an offset of $0.06\pm0.02$ dex between clumpy and non-clumpy galaxies in a comparable sample of 23 star-forming galaxies at $z\approx1.5$ using existing data from the FMOS-COSMOS survey. Similarly, lower {\nii}/{\halpha} ratio are typically found in galaxies that have more of their $\mathrm{UV_{rest}}$ luminosity originating from clumps, suggesting that \enquote{clumpier} galaxies are more metal poor. We also derive the intrinsic velocity dispersion and line-of-sight rotational velocity for galaxies from the GNIRS sample. The majority of galaxies have $\sigma_0/v_c \approx 0.2$, with no significant difference between clumpy and non-clumpy galaxies. Our result indicates that clump formation may be related to the inflow of metal-poor gas; however, the process that forms them does not necessarily require significant, long-term kinematic instability in the disk.
\end{abstract}



\section{Introduction}\label{sec:intro}
An important step to understanding the formation and evolution of galaxies is to trace where star formation is occurring within galaxies through cosmic time. Deep field observations reveal an abundance of high-redshift star-forming galaxies (SFGs) that host prominent, clumpy star formation \citep{Wuyts2012, Murata2014, Guo2015, Forster2018, Sok2022, Sattari2023}. Interestingly, the fraction of clumpy galaxies evolves with the cosmic star formation rate density, where it peaks at cosmic noon before declining toward higher redshift \citep{Shibuya2016}; however recent results from gravitationally-lensed galaxies observed with JWST suggest clumpy star formation may dominate the mass assembly of galaxies at $z > 4$ (e.g., \citealt{Vanzella2022, Adamo2024, Mowla2024}). Star-forming clumps are observed as UV-bright, kiloparsec-scale structures, but are also detected in {\halpha} emission. The specific star formation rate (sSFR) of clumps is elevated compared to the surrounding regions (\citealt{Wuyts2012, Wuyts2013}). A gradient in the sSFR of clumps is also reported by \cite{Guo2018}, where clumps that are further away from the galactic centers typically have higher sSFR. They found a similar negative gradient in stellar masses and ages, possibly suggesting an inward migration of clumps. In general, clumps have an estimated stellar masses of $10^7-10^9 ~\mathrm{M_\odot}$ (e.g., \citealt{Soto2017, Zanella2019, Huertas2020}), with lower mass estimates report in lensed galaxies \citep{Cava2018}. Nonetheless, the ubiquity and unique mode of clumpy star formation during a period known for galaxy mass assembly suggest that clumps play an integral role in galaxy formation. 


The two commonly accepted modes for the formation of star-forming clumps are (1) of \textit{in-situ} origins induced by violent disk instability (VDI), and (2) of \textit{ex-situ} origins induced by merger events. In the former case, clumps form due to the fragmentation of unstable regions within a gas-rich disk (e.g., \citealt{Agertz2009, Dekel2009a, Ceverino2010}), with instabilities predicted by the Toomre Q stability parameter \citep{Toomre1964}. Such gas-rich disks are suggested to be continuously fed by cosmological gas stream \citep{Dekel2009b}. Supporting observational evidence for clump formation via VDI comes from the analyses of the co-evolution of the clumpy fraction, merger rates and disk instability rates, which suggests that mergers alone cannot fully explain clump formation at $z\sim2$ (e.g., \citealt{Guo2015, Adams2022, Sattari2023}). Furthermore, \cite{Martin2023} argue that the higher star formation found in clumpy galaxies compared to non-clumpy galaxies is unlikely explained by mergers alone, and therefore must be due to some internal processes. Similarly, the observed stellar mass function of clumps is also consistent with clumps having an \textit{in-situ} origin (\citealt{Dessauges2018, Huertas2020}).

Although there is strong evidence for VDI-driven formation of clumps, clump formation via merger events is also supported by several studies. \cite{Guo2015} argue that minor mergers may drive clump formation in low-mass galaxies, whereas VDI dominates in high-mass galaxies. Merger-driven clump formation may also be more important at higher redshifts (e.g., $z>4$), as higher merger rates are observed at earlier times \citep{Duncan2019, Shibuya2022, Duan2024}. These merger-induced clumps are more commonly observed in simulations of major mergers at $z>6$ \citep{Nakazato2024}. 

Under the VDI framework, we postulate that if clumps form within a gas-rich disk that is replenished by (relatively) pristine gas from cosmological accretion, it is expected that the accreting gas would dilute the metal contents within the disk, leading to lower metallicities in clumpy galaxies compared to non-clumpy galaxies. Indeed, such a scenario has been observed in a simulation by \cite{Ceverino2016}, where clumps are found to have lower metallicities by 0.3 dex compared to the surrounding regions. Therefore, one way for testing \textit{in-situ} clumpy formation is to compare the metallicity of the gas disk for clumpy and non-clumpy galaxies (i.e., galaxies that do not host star-forming clumps).


In this work, we use the Gemini-North Near-Infrared Spectrograph (GNIRS) on the Gemini-North observatory to indirectly measure the gas-phase metallicity for a sample of clumpy and non-clumpy galaxies at $z\approx0.7$ and test the validity of clump formation under violent disk instability. We also make use of another large near-infrared spectroscopic survey (FMOS-COSMOS; \citealt{Silverman2015, Kashino2019}), which will serve as an independent test to see if the metallicity offset is observed in a different sample of SFGs at $z\approx1.5$. The metallicity is inferred from the {\nii} and {\halpha} line ratio, following the calibration from \cite{Pettini2004}. This line ratio is commonly used as both lines are easily accessible from the same filter, and their proximity mitigates the effect of dust in the inferred metallicity. The structure of the paper is as follows. First, we discuss our galaxy sample, as well as define the nomenclature for clumpy and non-clumpy galaxies in Section \ref{sec:sample}. Overviews of the GNIRS observations and data reduction are then reported in Section \ref{sec:data} and \ref{sec:spectra}, respectively. In Section \ref{sec:metallicity}, we present our metallicity measurements and compare them to other studies. Given the high signal-to-noise ratio of some of our spectra, we also investigate the relation between disk kinematics and morphologies (i.e., clumpy vs non-clumpy) in Section \ref{sec:kinematics}. Finally, we discuss the implications of our results in the context of clump formation due to violent disk instability in Section \ref{sec:discussion}. 

Throughout the paper, we adopt a $\Lambda$CDM cosmological model of the universe with $\Omega_\lambda$ = 0.7, $\Omega_\mathrm{M}$ = 0.3, and a Hubble constant of $H_0$ = 70 km/s/Mpc. All physical parameters are determined by assuming a Chabrier initial mass function \citep{Chabrier2003}, and our magnitudes are reported in AB magnitudes.

\section{Galaxy Sample} \label{sec:sample}

\subsection{GNIRS Sample}

Our targets are selected using the publicly available photometry of the COSMOS/UltraVISTA catalog \citep{Muzzin2013}. A number of criteria are employed to determine the candidates. We first require that the galaxies are at a redshift, either photometric or spectroscopic where available, where both {\halpha} and {\nii} lie within the wavelength coverage of the \textit{X} and \textit{J} filter of GNIRS, which avoid the highly contaminated atmospheric skylines in the \textit{H}-band. The choice in the two GNIRS filters correspond to the detection of {\halpha} and {\nii} between the redshift range of $0.6<z<0.9$. 

Using the stellar masses and SFRs from the UltraVISTA catalog, we then select galaxies of similar masses and SFRs to ensure that any observed metallicity offset is not driven by the fundamental metallicity relation. The masses were derived by fitting the spectral energy distribution of 30 photometric filters, extending from $0.15-24 ~\mathrm{\mu m}$, and the SFRs were determined from the rest-frame UV and infrared luminosity. We select primarily galaxies with $\log(\mathrm{M}_*/\mathrm{M_\odot}) \approx 10.5$. This corresponds to star-forming galaxies with a high SFR ($\mathrm{{\gtrsim}10 ~M_\odot yr^{-1}}$), and therefore increases the likelihood of a strong detection for both {\halpha} and {\nii}. Lastly, we select galaxies of similar colors (i.e., $U-V$ and $V-J$) to mitigate the effect of dust in our analyses as galaxies with higher dust attenuation generally have higher metallicities. 

\begin{figure}
    \centering
    \includegraphics[width=0.975\columnwidth]{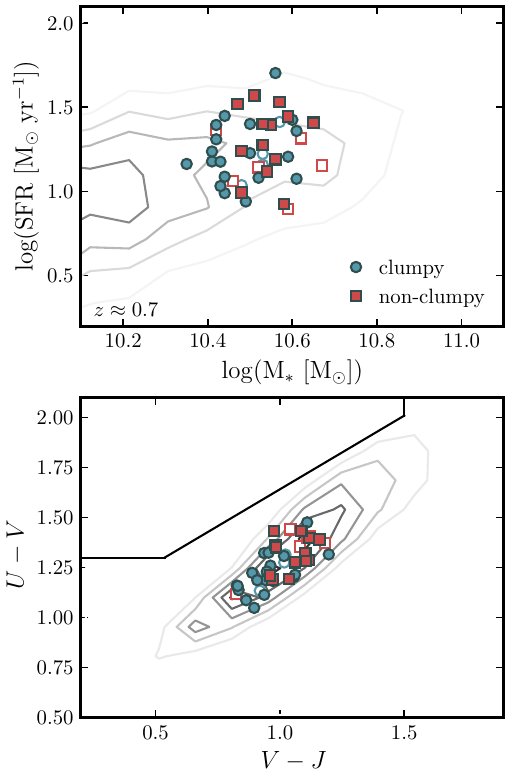}
    \caption{Top: The contour lines show the star-forming main sequence for the parent sample of galaxies at $z\approx0.7$ in the COSMOS field. The blue circle and red square markers denote clumpy and non-clumpy galaxies that were observed with GNIRS, respectively. The solid markers denote galaxies with detection in both {\halpha} and {\nii} (SNR$>3$).  Bottom: Similar to the top panel, but is now showing the \textit{UVJ} diagram. In general, the selection process is limited to galaxies of similar masses, SFRs and colors. This ensures that we can directly test for a relation between clumpy morphologies and metallicities. }
    \label{fig:sample_selection}
\end{figure}

Figure \ref{fig:sample_selection} summarizes our GNIRS galaxy sample. We select 42 galaxies; 23 are clumpy galaxies and 19 are non-clumpy galaxies (the classification of clumpy and non-clumpy is presented in Section \ref{sec:normprofile}). \edit1{The solid markers denote galaxies that have a detection of $\mathrm{SNR}>3$ in both {\halpha} and {\nii}. Throughout the analysis we use only these S/N $>$ 3 galaxies, which reduces our sample to 32 galaxies.} The average mass for the clumpy galaxies is $\mathrm{\log(M_*/M_\odot)} = 10.48^{+0.15}_{-0.02}$, with an average SFR of $\mathrm{\log(SFR ~[\mathrm{M_\odot~yr^{-1}}])} = 1.23^{+0.19}_{-0.14}$. The average mass for non-clumpy galaxies is $\mathrm{\log(M_*/M_\odot)} = 10.54^{+0.04}_{-0.06}$, and their average SFR is $\mathrm{\log(SFR ~[\mathrm{M_\odot~yr^{-1}}])} = 1.30^{+0.12}_{-0.28}$. The average specific SFR for both clumpy and non-clumpy galaxies is approximately $\log(\mathrm{sSFR ~[yr^{-1}]}) \approx -9.2$. Similarly, clumpy galaxies have an average $U-J = 1.21^{+0.11}_{-0.07}$ and $V-J = 0.95^{+0.11}_{-0.08}$, and for non-clumpy galaxies, $U-J = 1.32^{+0.08}_{-0.11}$ and $V-J = 1.05^{+0.06}_{-0.08}$. Here, the lower and upper errors are derived from the 16-th and 84-th percentiles.


\subsection{FMOS-COSMOS Sample} \label{subsec:fmos_selection}

The FMOS-COSMOS near-infrared spectroscopic survey \citep{Silverman2015, Kashino2019} utilized the Fiber-Multi Object Spectrograph on the Subaru Telescope, and comprised of an observing mode that primarily targets star-forming galaxies at $z\approx1.5$. In particular, the survey provides fluxes for both {\halpha} and {\nii}. The primary goal of this sample is to test whether a metallicity offset is also detected for a completely different sample of galaxies that have been observed with a different observing strategy, but are selected similarly in masses, SFRs, and colors as the GNIRS sample.

We cross-match galaxies from the FMOS catalog to the UltraVISTA catalog, keeping only galaxies that have a measured RA and Dec that are within 0.2 arcsec of both catalogs. This angular scale corresponds to less than two imaging pixels of the UltraVISTA/COSMOS mosaics (which has a pixel scale of 0.15 arcsec). This is also well below the size of the images' point spread function, and therefore ensures that there is no contamination from close-by neighbours. We select FMOS galaxies that have {\halpha} and {\nii} fluxes detected to at least 3 sigma. Similar to the GNIRS sample, we also select galaxies within a mass range between $10.4 < \mathrm{\log(M_*/M_\odot)} < 10.6$, and limit the analyses to galaxies of similar colors, i.e., $0.9<U-V<1.2$ and $0.7<V-J<1.1$. In total, there are 23 FMOS galaxies at $z\approx1.5$ that meet our selection criteria. We reiterate that the goal for this sample is mainly to determine whether a metallicity offset is similarly observed between clumpy and non-clumpy galaxies when using a different sample of galaxies. In the following sections, we discuss how galaxies are classified as clumpy and non-clumpy from their ground-based images.

\subsection{Resolving clumps with finite-resolution deconvolution}

While clumpy galaxies are ubiquitous at high-redshift, identifying galaxies as clumpy is difficult due to the fact that high-resolution observations are required to resolve the kiloparsec-scale structures. Furthermore, multi-wavelength images are typically needed to account for the morphological K-correction effect, in which the morphology of galaxies may change depending on which wavelength they are observed in. This becomes particularly important when the galaxies are observed over a wide range of redshifts. 

A unique solution to obtaining high-resolution and multi-wavelength imaging for a large volume of the sky is to use image deconvolution. This had been previously done in \cite{Sok2022} using finite-resolution deconvolution (\texttt{FIREDEC}; \citealt{Cantale2016a}), where we deconvolved the multi-band imaging of approximately 20,000 SFGs in the COSMOS field at $0.5<z<2$, and created resolved stellar mass and surface brightness maps for those galaxies. That work showed that \texttt{FIREDEC} can recover clumps in galaxies from ground-based imaging. Comparing the deconvolved images to HST observations confirmed that these clumpy features are intrinsic to the galaxies and not artifacts of deconvolution. 

A technical overview of image deconvolution and \texttt{FIREDEC} is provided in \cite{Cantale2016a}, with practical applications discussed in \cite{Cantale2016b} and \cite{Sok2022}. In summary, this deconvolution algorithm aims to to deconvolve images to a finite resolution, ensuring that the resulting image retains its own PSF (i.e., as opposed to completely removing the PSF to achieve an infinite angular resolution, which would violate the Nyquist's sampling theorem). In practice, a target resolution is defined to be a 2D Gaussian function. The deconvolution kernel is then constructed through a combination of analytical and numerical fits, such that the convolution between the deconvolution kernel and the Gaussian functions gives back the PSF of the image. 

\begin{figure*}
    \centering
    \includegraphics[width=\textwidth]{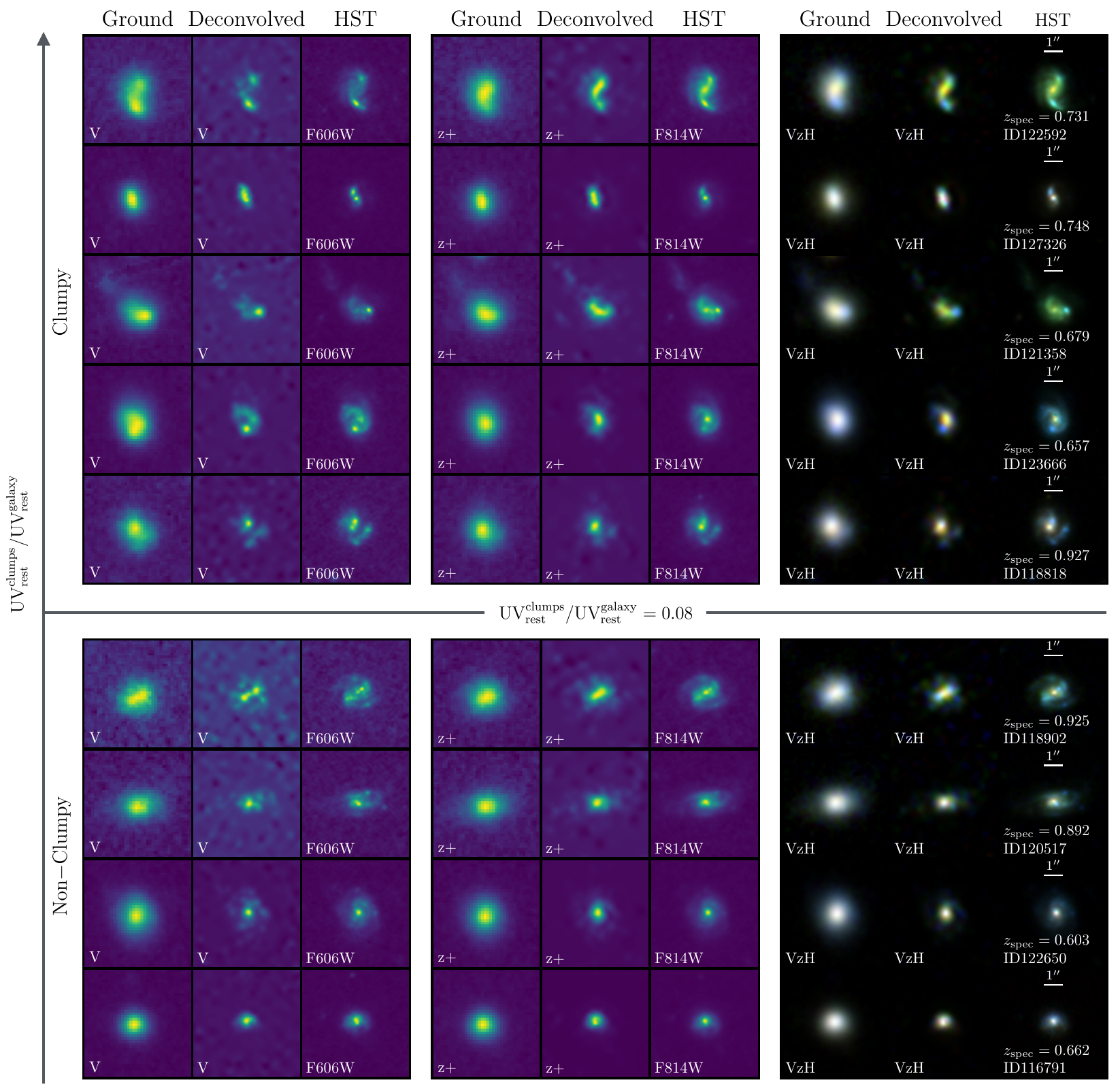}
    \caption{Comparison of deconvolved images to the corresponding HST images. From the left to right panel, we compare the \textit{V}-band imaging to \textit{F606W}, \textit{z+}-band imaging to \textit{F814W}, and the composite \textit{VzH} image to a composite image constructed from \textit{F606W}+\textit{F814W}+\textit{F160W}. Each stamp is $7.8^{\prime\prime} \times 7.8^{\prime\prime}$. For each panel, the left to right column shows the ground-based image, the deconvolved image, and the corresponding HST image. Only 9 out of the 42 galaxies from our sample have HST imaging in the 3 HST filters. The images are deconvolved to a resolution of 0.3", corresponding to a physical scale of ${\sim}2.3 ~\mathrm{kpc}$ at $z\approx0.7$.  We arrange the figure such that the galaxies are shown in ascending order based on their clumpiness, as defined by the ratio of rest-frame UV light emitted from clumps relative to the host galaxy. Clumpy galaxies are those with clumps that have at least 8\% of the total rest-frame UV luminosity. }
    \label{fig:deconv_examples}
\end{figure*}

In order to deconvolve images in different filters of the COSMOS field, we constructed deconvolution kernels for each filter. This is required as the PSF can vary from filter to filter, which arises due to the different observing conditions (see \citealt{Sok2022} for the PSF variation for each filter of the COSMOS field). Here, we deconvolved to an angular resolution of 0.3" which probes a physical scale of approximately 2.3 kpc for galaxies at $z\approx0.7$. While this angular resolution only partially resolves star-forming clumps, it is sufficient for identifying whether a galaxy is clumpy or non-clumpy. A comparison between the ground-based imaging and the deconvolved images for a few galaxies in this study is shown in Figure \ref{fig:deconv_examples}. These galaxies are chosen as examples because they have multi-band HST imaging, which enables us to directly compare the deconvolved images to HST images. Galaxies that are classified as clumpy are denoted by the asterisk in the ID name in the right panel. In general, deconvolution reveals multiple clumps that are not detectable with ground-based imaging. These clumps are also detected by HST, suggesting that the resolved clumps are intrinsic to the galaxies, and not deconvolution artifacts. 

\subsection{Clumpy/Non-clumpy SFGs Classification} \label{sec:normprofile}

A number of methods have been used to classify whether a galaxy is clumpy or non-clumpy. Clumps can be identified by creating a contrast image, in which an image is passed through a high-pass filter (e.g., \citealt{Zanella2019, Claeyssens2023}). The flux and size of the clumps can be obtained by fitting each clump with a  Gaussian model. Recent studies have also employed a convolutional neural network to identify the location of clumps from an image \citep{Huertas2020}. 

In this work, we follow the method described in \cite{Wuyts2012}, which utilizes the normalized rest-frame \textit{U}-band light profile to identify clumpy regions in a galaxy. The rationale of a normalized profile relies on the fact that regions with relatively high surface brightness, such as UV-bright clumps, will appear as a bump in an otherwise smooth light profile of a galaxy. We will briefly describe the steps to making a normalized light profile, however we refer the reader to \cite{Wuyts2012} for a more detailed process. We first obtain $U_\mathrm{rest}$ by spatially fitting the multi-band deconvolved images with \texttt{EAZY} \citep{Brammer2008}. Instead of fitting the spectral energy distribution (SED) over individual pixels, which can introduce biases for pixels with low signal-to-noise ratio (SNR), we instead bin pixels using a Voronoi binning technique (\texttt{vorbin}; \citealt{Cappellari2003}) so that each Voronoi bin has a minimum SNR of 5 in the \textit{Ks}-band. A similar binning technique has been employed to construct stellar mass maps from spatially SED fits \citep{Wuyts2012, Sok2022, Tan2022}.

\begin{figure}
    \centering
    \includegraphics[width=0.975\columnwidth]{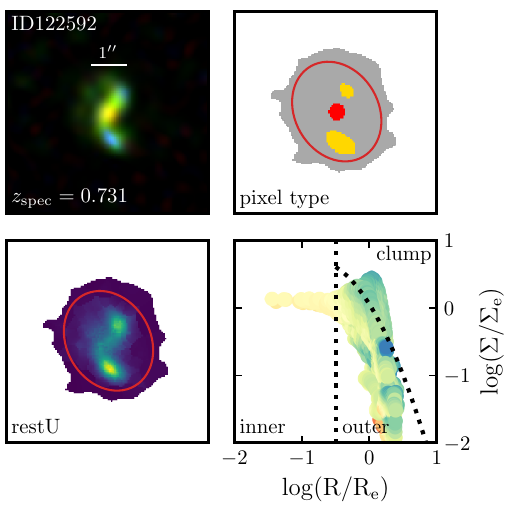}
    \includegraphics[width=0.975\columnwidth]{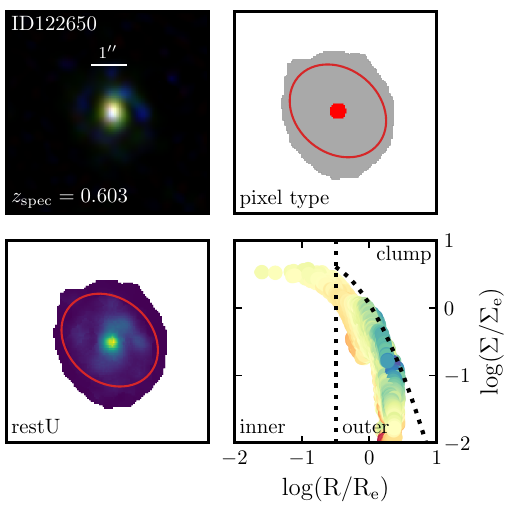}
    \caption{Two examples showing how the normalized light profile is used to identify clumps. The top figure shows the normalized profile for a clumpy galaxy while the bottom figure shows the profile for a non-clumpy galaxy. In each figure, the top left panel shows the composite \textit{VzH} image ($6^{\prime\prime}\times6^{\prime\prime}$), the bottom left shows the surface brightness, the bottom right shows the constructed normalized \textit{U$_\mathrm{rest}$} profile, and top left shows the pixel type map based on the division lines in the normalized light profile. The clump regime is defined as the pixel residing above the curved dotted line. As shown in the top figure in the pixel type map, clumps are identified easily using the normalized profile. The color-coding in the normalized profile respond to the $(U-V)_\mathrm{rest}$. Pixels in the clumpy regime are UV-bright.}
    \label{fig:normprofile}
\end{figure}

Once a map of the $U_\mathrm{rest}$ surface brightness is created, the normalized light profile for a galaxy is constructed. We characterize the galaxies by two elliptical parameters; the ellipticity and position angle. This is done by fitting an elliptical function to a segmentation map of the galaxy. These parameters are then used to define elliptical apertures centered on the galaxy. We then define the effective radius of the galaxy as the radius at which half of the light from the galaxy is within that radius. This is measured from a curve of growth based on the elliptical apertures. The effective surface brightness is then measured as the mean surface brightness within the effective radius. The normalized profile is constructed by normalizing the $U_\mathrm{rest}$ surface brightness at each pixel by the effective surface brightness, and the galactocentric distance of each pixel by the effective radius. In essence, the normalized profile illustrates how the normalized surface brightness varies with distance from the center of the galaxy. Figure \ref{fig:normprofile} shows two examples of the normalized light profile for a clumpy and non-clumpy galaxy. 

Clumps are situated in the same parameter space within the normalized profile. The division lines of the inner, outer, and clumpy regions, shown in Figure \ref{fig:normprofile}, are defined in Eqn. 6-8 of \cite{Wuyts2012}, based on their visual inspections of the stacked profiles of their sample. Galaxies are defined as clumpy if they have pixels that originate from the clumpy regime in the normalized profile, and that contribute ${\geq}8\%$ of the total $U_\mathrm{rest}$ luminosity of the host galaxy. This criterion is commonly used to identify star-forming clumps in high-redshift galaxies in other studies, and therefore represents a consistent way of classifying clumpy galaxies. 

\section{GNIRS Observations} \label{sec:data}
A total of 42 galaxies in the COSMOS field were observed with GNIRS on the Gemini-North observatory. These observations took places between 2019 and 2022 as part of several GNIRS observation programs, with program IDs: GN-2019B-Q-131, GN-2019B-Q-229, GN-2020A-Q-307, GN-2020B-Q-215, GN-2021B-Q-212, GN-2022A-Q-213. The observations were done through the long-slit spectroscopy mode in the \textit{X} or \textit{J} filter and the 111 line per mm grating. We use the short camera, corresponding to a pixel resolution of 0.15 arcsec per pixel, with a slit opening of 1 arcsec. The observational setup therefore corresponds to a wavelength coverage between 1.03 - 1.37 microns, at a spectral resolution of $R\approx2000$ ($\sigma_{\mathrm{inst}}\sim50~\mathrm{km/s}$). This spectral resolution is sufficient for differentiating the {\halpha} and {\nii} emission lines. The on-source integration time for each galaxy is set either to 40, or 60 minutes. The shorter integration time is used when the observations took place with good seeing (i.e., Band 1) condition, while the longer 60 min of integration is needed when observing in the degraded seeing of Band 3 to account for greater light loss through the 1" wide slit opening.

A blind offset from a reference bright star (or a reference object when there are no nearby star) is needed in order to center the slit onto our faint galaxies. The position of the slit is fixed along the major axis of the galaxy to ensure that most of the light coming from the galaxy is captured by the slit. We observe each galaxy using the ABB$^\prime$A$^\prime$ pattern, which corresponds to the dithering nod of -1", 6", -6", 1" along the slit direction. This pattern is used for sky subtraction, but helps mitigate pixel persistence when observing bright objects such as a standard star. At each nodding position, we have 300 seconds of integration. The standard star is observed either before, or after the exposures on the galaxy using the same observational setup. The spectroscopic observations of a standard star is needed in order to apply flux calibration for telluric absorption features in the near-infrared. 

\section{Near-Infrared Spectra} \label{sec:spectra}
The reduction of the raw data is done using \texttt{Gemini IRAF} \citep{GeminiIRAF2016}. A number of raw exposure frames for the science observations were affected by striping patterns; however these patterns were removed with \texttt{cleanir}\footnote{https://github.com/andrewwstephens/cleanir}. Telluric and science observations are then flat-fielded, sky-subtracted and wavelength-calibrated. The master flat is obtained by stacking flat frames that do not have anomalous mean, or standard deviation. The wavelength solution is computed by running \texttt{nswavelength} on the included Argon lamp observations. We extract the one-dimensional spectrum for the standard star using \texttt{nextract}. In the following sections, we describe the steps to extract the one-dimensional spectra from the reduced two-dimensional science spectra.

\subsection{Spectra and velocity curves} \label{sec:spectex}

One simple method for extracting a one-dimensional spectrum from the two-dimensional spectrum is the boxcar method, in which the two-dimensional spectrum is summed along the slit direction. However, this can lower the SNR of the emission lines if the galaxy's rotation is not accounted for. In this section, we describe the steps to extract the one-dimension spectrum by modeling the kinematics of the galaxy. 

\begin{figure*}[ht!]
    \centering
    \includegraphics[width=\textwidth]{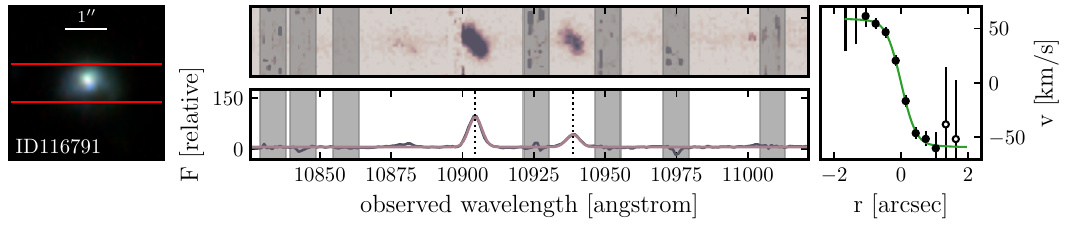}
    \includegraphics[width=\textwidth]{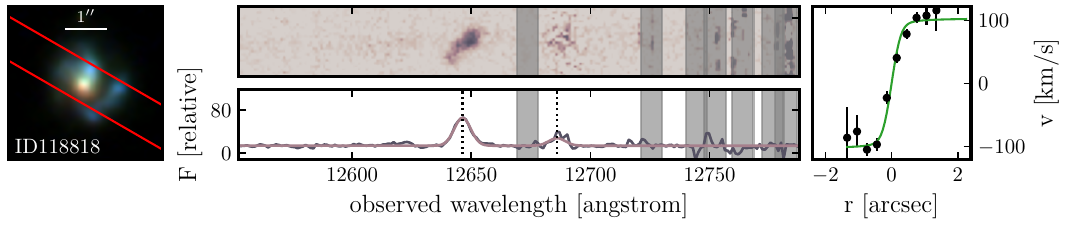}
    \includegraphics[width=\textwidth]{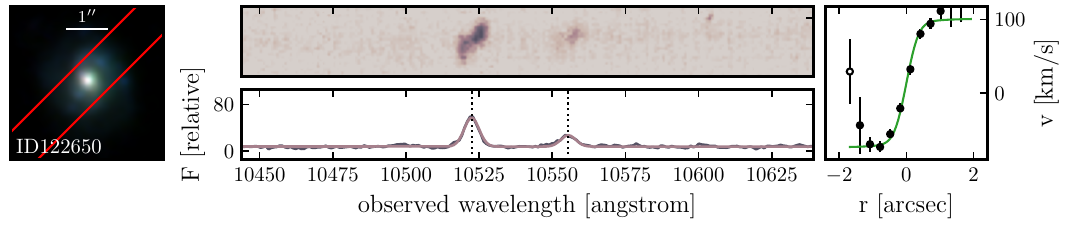}
    \includegraphics[width=\textwidth]{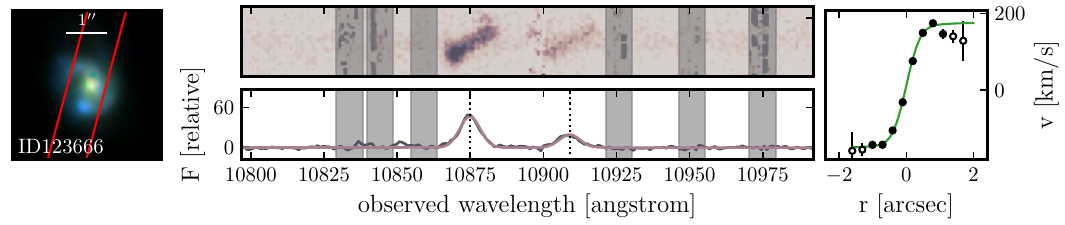}
    \caption{Examples of spectra along with the rotation curve of the galaxies. The left panel shows the composite image of the galaxies. The red line denotes the slit, which is purposely chosen to lie along the axis containing clumps. The top middle panel shows the two-dimensional spectrum where the y-axis is the slit (or spatial) direction and the x-axis is the dispersion direction. The bottom middle panel shows the extracted 1D spectrum after correcting for the rotation curve. The grey highlights are regions that are contaminated by skylines. The right panel shows the rotation curve of the galaxies, where the green line is the fitted rotation curve, and the black markers are the measured centroids of the emission line at each spatial bin along the two-dimensional spectrum.}
    \label{fig:reduced_1d_spectra}
\end{figure*}

The kinematics modeling is performed using similar steps as outlined in \cite{Weiner2006}. We first obtain the velocity and dispersion profile of a galaxy by fitting a Gaussian function to its emission lines along the slit. To this end, we bin the two-dimensional spectrum along the spatial direction by 2 pixels to increase the SNR of the emission lines. Here, {\halpha} is primarily used when fitting the Gaussian function as it has a higher SNR compared to {\nii}. However, {\nii} is used when {\halpha} lies next to a strong skyline as the centroiding of the Gaussian function may be affected by the skyline. The fitted parameters are determined through a least-squares Levenberg-Marquardt minimization algorithm with respect to the observed data using \texttt{lmfit} \citep{Newville2016}. The retrieved error represents an uncertainty of one standard deviation around the fitted values. A few examples of the measured velocity profile are plotted on the right panels of Figure \ref{fig:reduced_1d_spectra}. 

The kinematics model consists of two components; a rotation curve and the dispersion term. We assume that the intrinsic velocity dispersion $\sigma_0$ is constant with radius for each galaxy. The inverse tangent function is adopted for the rotation of the galaxies, 
\begin{equation}\label{eqn:rotation}
    v(r) = v_0 + \frac{2}{\pi} v_a \arctan\Big( \frac{r - r_0}{r_t} \Big), 
\end{equation}

\noindent where $v_a$ is asymptotic rotation velocity, $r_o$ is the spatial center of the galaxy, and $r_t$ is the knee radius to the flat part of the curve. When modeling the emission line, we vary values for $v_0$, $r_0$, $v_a$ and $\sigma_0$, while keeping $r_t$ constant at a fiducial value of $0.2^{\prime\prime}$ due to the limited seeing. The modeled emission line is then spatially convolved with a Gaussian seeing kernel. We determine the appropriate FWHM from the acquisition images of the bright offset star, or from the acquisition images of the standard star (which are observed before or after the science integration). The modeled emission line is also convolved by the line spread function. We use \texttt{emcee} \citep{Foreman2013} to explore the parameter space, assuming a flat prior for each parameter. The best-fit parameters and their errors are taken to be the median and standard deviation, respectively. We note that $\sigma_0$ is not resolved for a few galaxies due to our relatively coarse spectral resolution of $\mathrm{R}\sim2000$.

The rotation models are then used to align the emission lines along the slit. The one-dimensional, velocity-corrected spectrum are then optimally extracted, using the extraction algorithm from \cite{Horne1986}. The weighting profile is determined from the spatial profile of the emission lines. In cases where the emission lines have low SNR, we perform a standard boxcar extraction over some aperture width (i.e., 2.1$^{\prime\prime}$ to 2.7$^{\prime\prime}$ depending on the size of the galaxy). Finally, we do not apply an aperture correction for possible slit losses. While the size of galaxies varies and can therefore affect the measured {\nii}/{\halpha} ratio, we mitigate this by placing the slit along the major axis that contains clumps. A slit correction would also assume some generalization of the metallicity gradient, which can be diverse (e.g., \citealt{Gillman2021}). However, larger galaxies typically have a negative metallicity gradient \citep{Carton2018}. Since clumpy galaxies are generally larger compared to non-clumpy galaxies (e.g., \citealt{Martin2023}), and assuming a negative metallicity gradient for clumpy galaxies, we expect the measured metallicity for clumpy galaxies to be overestimated compared to non-clumpy galaxies when using a slit that is centered on the central region of the galaxies. Given these reasons, we choose to not apply an uncertain slit-loss correction and present the data as-is. Finally, we note that the size difference between non-clumpy and clumpy galaxies is observed for the GNIRS sample (e.g., see Fig. \ref{fig:n_reff_vs_ufrac}), where we find the average effective radius of clumpy and non-clumpy galaxies to be $5.1\pm0.3$ and $3.4\pm0.3$ kpc (corresponding to $0.7^{\prime\prime}$ and $0.57^{\prime\prime}$), respectively. Figure \ref{fig:reduced_1d_spectra} summarizes this section, showing examples of the observed rotation in the two-dimensional spectrum, the derived velocity curve, and the extracted spectrum after correcting for the velocity curve.

\subsection{Telluric corrections}

Molecules in the atmosphere can absorb and subsequent suppress certain spectral features in the near- and mid-infrared. These telluric absorption features may therefore affect the ratio between {\nii} and {\halpha} if the two lines fall within regions of strong absorption. Correcting for the absorption features is done by observing the spectrum of a standard star. Each of our observing blocks included a near-infrared spectroscopic observation of a standard star. The standard stars are selected to be early-type stars at a similar airmass, and are selected primarily because their spectra are generally devoid of intrinsic stellar absorption features in the near-infrared (between $1.05-1.3$ microns), with the exception of hydrogen recombination lines. However, these hydrogen lines can be fitted and removed. The steps to apply the telluric correction to our observations are listed below. 

\begin{enumerate}
    \item Hydrogen absorption lines in the standard star telluric observations are first masked out. The continuum of the masked spectrum is modeled as the sum of Chebyshev polynomials using \texttt{specutils}. 
    \item The reference spectrum for each standard star is obtained from the stellar spectral library of \cite{Pickles1998}. These reference spectra are then flux-calibrated so that their \textit{V}-band magnitude is equal to the observed magnitude of the respective star. We mask the hydrogen lines in the flux-calibrated reference spectra and model their continuum using a similar step as above. 
    \item The correction factor at each wavelength is obtained by dividing the flux-calibrated spectrum by the observed telluric spectrum.
\end{enumerate}

\noindent In general, we find that the correction factor for the ratio of {\nii} and {\halpha} ranges between 0.8 and 1.5. On average, the correction factor for the clumpy and non-clumpy sample is 1.05 and 1.09, respectively. 

\section{Comparison between Clumpy \& Non-Clumpy Galaxies} \label{sec:metallicity}
\subsection{{\nii}/{\halpha}}

Figure \ref{fig:reduced_1d_spectra} shows four examples of the reduced one-dimensional, telluric-corrected spectrum. The flux for each emission line is obtained by fitting a double-Gaussian function using \texttt{lmfit}. To this end, we fix the relative wavelength offset between the two emission lines, and assume that the line width for both lines are the same. Therefore, our fitting function consists of five parameters; the peak intensity ($a$) for both {\nii} and {\halpha}, the line width ($\sigma$) for the emission lines, the wavelength center for {\halpha}, and a constant continuum/background noise. The integrated flux for both emission lines is calculated as $a\sigma\sqrt{2\pi}$, using the respective line intensity. Similarly, the flux error is obtained by propagating the uncertainties of the fitted parameters $a$ and $\sigma$. 

Out of the 42 galaxies, 2 galaxies are not detected in either {\halpha} and {\nii}. Only 32 out of the remaining 40 galaxies have {\halpha} and {\nii} lines that are not significantly affected by atmospheric skylines, and have an SNR $> 3$. Table \ref{tab:gnirs} lists their measured line ratios and their associated error. We find the weighted average of the {\nii}/{\halpha} ratio for the entire sample to be $0.36\pm0.01$. When separated into clumpy and non-clumpy sample, we find that clumpy galaxies have a weighted flux ratio average of $0.31\pm0.02$, while non-clumpy galaxies have a flux ratio of $0.42\pm0.02$. The difference in the {\nii} and {\halpha} flux ratio is $0.11\pm0.02$. Taken at face value, this suggests a clear metallicity difference between the two samples; however, in the next section we examine this difference more closely via the mass-metallicity relation and the fundamental metallicity relation.

\subsection{The mass-metallicity relation}

The gas-phase metallicity of each galaxy is inferred by converting the {\nii}/{\halpha} line ratio to an oxygen abundance, using the N2 line ratio calibration from \cite{Pettini2004}, which has the following form, 

\begin{equation}
    12 + \mathrm{\log(O/H)} = 8.9 + 0.57 \times \log(\mathrm{[NII]\lambda6584}/\mathrm{H\alpha})
\end{equation}

\noindent This calibration is valid for $-2.5<\log(\mathrm{[NII]}/\mathrm{H\alpha})<-0.3$. The error associated with the slope and intercept are 0.03 and 0.04, respectively. The {\nii}/{\halpha} line ratio and inferred metallicities are listed in Table \ref{tab:gnirs}.

\begin{figure*}
    \centering
    \includegraphics[width=\textwidth]{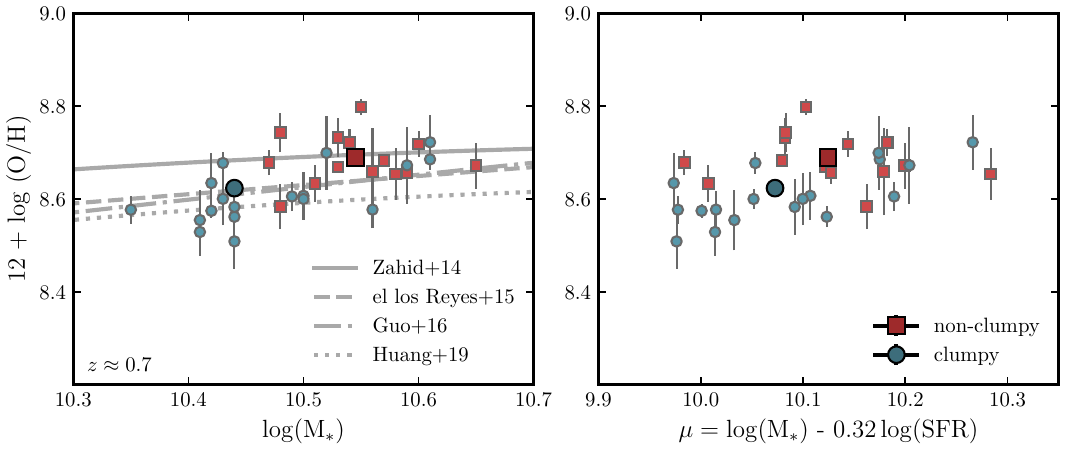}
    \caption{Left: the metallicities of clumpy and non-clumpy galaxies in relation to stellar masses. The dark, larger markers show the weighted average. The different lines show the mass-metallicity relation for galaxies at $z\approx0.8$ from \cite{Zahid2014, Reyes2015, Guo2016}, and \cite{Huang2019}. The error is calculated as the weighted standard error. Right: The metallicities in relation to the projected stellar masses and star formation rate.}
    \label{fig:gnirs_metallicity}
\end{figure*}

The left side of Figure \ref{fig:gnirs_metallicity} shows the mass-metallicity relation for 32 SFGs with a detected SNR of at least 3 for both {\halpha} or {\nii}, color-coded as clumpy vs non-clumpy. The weighted averages are calculated using the inverse variance weighting, and are denoted as the dark, large markers, and their errors are calculated as the weighted standard error. We also plot the mass-metallicity relation from other studies for galaxies at a similar redshift range. However, since these studies used a different metallicity calibration, which can lead to discrepancies in metallicities (e.g., \citealt{Kewley2008}), we also apply a polynomial correction to each mass-metallicity relation to convert them to the N2 calibration. The polynomial correction are taken from \cite{Teimoorinia2021}.

The grey curves are the corrected mass-metallicity relation from \cite{Zahid2014}, \cite{Reyes2015}, \cite{Guo2016}, and \cite{Huang2019}, respectively. In general, we find that our values are consistent with other studies. Massive galaxies typically have higher metallicities, while lower-mass systems having lower metallicities. When split into a clumpy and non-clumpy sample, we find that the sample of clumpy SFGs have a lower median metallicity, with an offset of $0.07\pm0.02$ dex compared to their non-clumpy counterparts. 

While the mass-metallicity relation of Figure \ref{fig:gnirs_metallicity} suggests a metallicity offset between clumpy galaxies and non-clumpy galaxies, we acknowledge that there is a small (but non-negligible) mass offset between the two samples. Our clumpy sample has a lower mean stellar mass compared to the non-clumpy galaxy sample. Although the original sample of targets were chosen so that both the stellar masses and star formation rates are similar, some galaxies were affected by skylines. This resulted in two samples of galaxies with a mean mass offset of $0.1$ dex. However, we do not expect the mass difference to affect the observed metallicity offset by much since the mass-metallicity relations suggest only a slight change in metallicities with our given mass offset. For example, given the mass offset of 0.1 dex, the metallicity is expected to change by 0.01-0.02 based on the different mass-metallicity relations. 

\begin{figure}
    \centering
    \includegraphics[width=\columnwidth]{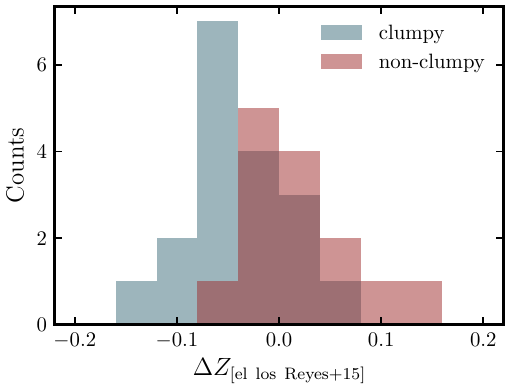}
    \caption{The histogram of the differences between the observed metallicity and the re-normalized mass-metallicity relation from \cite{Reyes2015} for clumpy and non-clumpy galaxies. Using a KS test, we find the two distributions are statistically different, with a p-value of 0.004. }
    \label{fig:delta_z}
\end{figure}

We further quantify this by fitting a mass-metallicity relation to our data. Using the analytical functions from \cite{Zahid2014}, \cite{Reyes2015}, \cite{Guo2016}, and \cite{Huang2019}, we find the best fitting mass-metallicity relation for our observed metallicities, by fixing the slope and only fitting the y-intercept. We opt to do this instead of fitting for both the slope and normalization factor due to our limited sample size and mass range. We find that the mass-metallicity relation from \cite{Reyes2015} gives that smallest scatter in the residuals. Figure \ref{fig:delta_z} shows the histogram of the differences between the metallicity of clumpy and non-clumpy galaxies to the fitted mass-metallicity relation. A two-sample Kolmogorov-Smirnov (KS) test is performed on the two distributions to determine whether the distributions are statistically similar ($p>0.05$), or different ($p\leq0.05$). We obtain a p-value of 0.004, which rejects the null hypothesis of the same distribution.

Finally, we note that the slope of the N2 relation from \cite{Pettini2004} is shallower (particularly at higher N2 regime - $\log(\mathrm{N2})\sim~0.5$) relative to other N2 calibrations (e.g., \citealt{Marino2013}). This could lead to a more defined metallicity offset. To test whether the observed metallicity offset is affected by the choice of line calibration, we perform a similar KS test as above, but using metallicities derived from the steeper line calibration of \cite{Marino2013}. Here, We obtain a p-value of 0.03. While the result is less significant than before, the null hypothesis that clumpy galaxies have the metallicities compared to non-clumpy galaxies can still be rejected.

\subsection{The fundamental metallicity relation}

It is known that the gas-phase metallicity has a secondary dependence on the star formation rate of the galaxies. The metallicity is found to decrease with increasing SFR at a given stellar mass. The combined stellar mass, SFR and metallicity is suggested to form a fundamental plane, and is known as the fundamental metallicity relation (FMR; \citealt{Mannucci2010}). We use the following definition from \cite{Mannucci2010} to project the three-dimensional parameter space into two dimensions, 

\begin{equation}
    \mu = \log(M_*) - 0.32 \log(\mathrm{SFR}), 
\end{equation}

\noindent This projection of the FMR is chosen as it minimizes the scatter of the gas metallicity. 

The right side of Figure \ref{fig:gnirs_metallicity} shows the projected FMR for our galaxies. As shown in the figure, the distribution of $\mu$ for the clumpy and non-clumpy sample is slightly offset, with a mean difference of $\Delta \mu = 0.04$. When comparing the two distributions of $\mu$, a $p$-value of 0.21 is obtained, suggesting that the distribution of $\mu$ (i.e., the stellar masses and SFRs) for clumpy and non-clumpy galaxies are drawn from the same parent sample. However, if we perform a KS test on the metallicity distributions for clumpy and non-clumpy galaxies, we obtain a $p$-value of 0.004. These results suggest that lower metallicities in clumpy galaxies are not driven by the mass of the galaxies and/or their SFR, and are therefore intrinsic to the clumpy population. 

\subsection{Metallicity Difference in FMOS-COSMOS} \label{sec:fmos_z_offset}
In this section, we investigate whether there is a metallicity difference between clumpy and non-clumpy galaxies at a higher redshift. We use the {\nii} and {\halpha} observations from the FMOS-COSMOS survey, which targets galaxies at $z\approx1.5$. 

The selection criteria was described in Section \ref{subsec:fmos_selection}. In essence, we select star-forming galaxies of similar masses, SFRs and colors, with $\mathrm{\log(sSFR ~[yr^{-1}])} = -8.70$ and -8.67 for clumpy and non-clumpy galaxies, respectively. Furthermore, only galaxies with a robust detection in {\halpha} and {\nii} are selected ($\mathrm{SNR}>3$). As these galaxies are located in the COSMOS field, we use the deconvolved dataset from \cite{Sok2022} to identify clumpy and non-clumpy galaxies (see Section \ref{sec:normprofile}). This results in a final sample size of 23 galaxies. 

\begin{figure}
    \centering
    \includegraphics[width=\columnwidth]{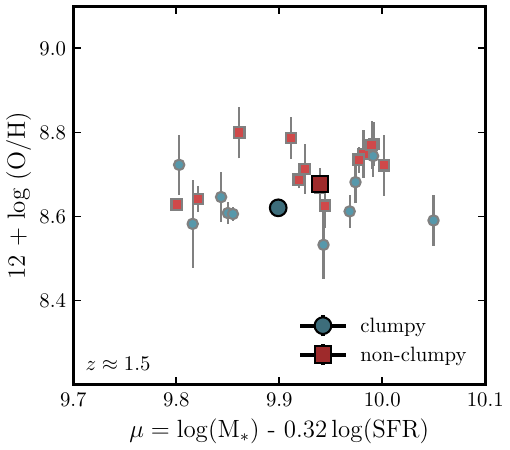}
    \caption{The fundamental metallicity relation of clumpy and non-clumpy galaxies in the FMOS-COSMOS sample. The darker markers shows the weighted average, and the error is taken as the weighted standard error. In general, the metallicity of clumpy and non-clumpy galaxies in the FMOS-COSMOS sample are similar to the GNIRS sample. The metallicity offset is measured to be $0.06\pm0.02$ dex.}
    \label{fig:fmos_fmr}
\end{figure}

Figure \ref{fig:fmos_fmr} shows the relation between metallicities and the projected stellar masses and SFRs for clumpy and non-clumpy galaxies from the FMOS sample. In general, we find that the metallicities of the FMOS sample are comparable to our values. This is also in agreement with other studies of the FMR, which indicated that there is no observed evolution in the FMR with redshift \citep{Sanders2021}. We also find that clumpy galaxies have a lower median metallicity compared to non-clumpy galaxies, with an offset of $0.06\pm0.02$ dex. A similar two-sample KS test was performed on the metallicity distribution for clumpy and non-clumpy galaxies, and we obtain a p-value of 0.02. Given that the metallicity offset is observed at two different redshifts for a selected sample of galaxies with similar stellar masses and SFRs, these results suggest that clumpy galaxies intrinsically have lower metallicities compared to non-clumpy galaxies, and this differences cannot be explained by differences in their stellar mass, or SFR alone. 

\section{Kinematics of Clumpy Galaxies} \label{sec:kinematics}

\begin{figure}
    \centering
    \includegraphics[width=\columnwidth]{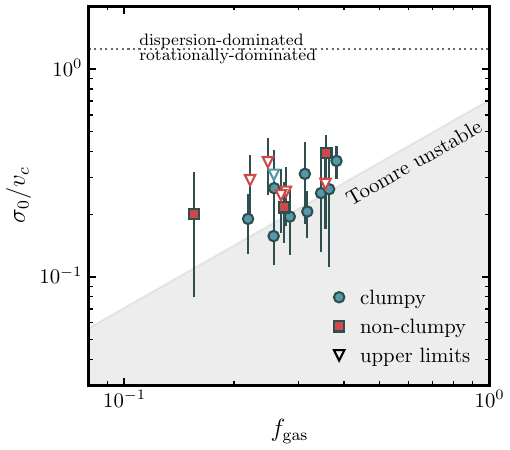}
    \caption{The relation between $\sigma_0/v_c$ and $f_\mathrm{gas}$ for clumpy and non-clumpy galaxies. $f_\mathrm{gas}$ is derived from the integrated SFR, using the Kennicutt-Schmidt relation. The darker color markers show the mean value, and the error is calculated as the standard error of the mean. The dotted line shows the typical cutoff between rotationally- and dispersion-dominated systems \cite{Forster2009}. The shaded region shows the parameter space in which the Toomre Q value is less than 1 (i.e., unstable disks). This relation is defined by Eqn. \ref{eqn:stable_disk} (see text for the derivation). Based on the integrated values, clumpy and non-clumpy galaxies have stable disks, with no significant differences between their kinematics.}
    \label{fig:kinematics_integrated}
\end{figure}

Having established that clumpy and non-clumpy galaxies have different metallicities, we now explore if additional signatures between the two samples are observed.  Kinematics are interesting as we might expect to see signs of increased disk instability in clumpy galaxies if clumps are formed via VDI. Figure \ref{fig:kinematics_integrated} shows the relation between the gas fraction $f_\mathrm{gas}$ and $\sigma_\mathrm{0}/v_c$. Here, $v_c$ is the asymptotic rotational velocity, corrected for the inclination of the galaxy (i.e., $v_c = v_a/\sin{i}$). The inclination $i$ is estimated as,

\begin{equation}
    \cos i = \sqrt{\frac{q^2 - q_0^2}{1 - q_0^2}}.
\end{equation}

\noindent $q$ is the axial ratio of the galaxy, and $q_0$ is the intrinsic axial ratio for an edge-on galaxy, taken to be $q_0=0.2$ \citep{Ryden2006}.  The gas fraction is inferred using the relation between gas and SFR density (e.g., the Kennicutt-Schmidt relation; \citealt{Kennicutt1998}). The SFR density is derived by assuming that star formation is contained within the effective radius. We obtain the effective radius and $q$ by fitting a S\'ersic profile to the \textit{F814W} images using \texttt{statmorph} \citep{Rodriguez2019}. We further limit the analyses to ``kinematically-resolved" galaxies, where $v_c/\delta v_c$ is greater than 1 ($\delta v_c$ is the measurement uncertainty). We also note that some of the derived values $\sigma_0$ are unreliable due to our relatively coarse spectral resolution. In cases where the observed velocity dispersion is within one sigma uncertainty of the instrumental dispersion, we plot the observed dispersion (i.e., $\sigma_\mathrm{obs} = \sqrt{\sigma_0^2 + \sigma_\mathrm{inst}^2}$) as the upper limit for the intrinsic dispersion. These are denoted as open markers in the figure. 

The dotted horizontal line in Figure \ref{fig:kinematics_integrated} denotes the cutoff between dispersion- and rotationally-dominated disks, with the former having $\sigma_0/v_c > 1.25$ (e.g., \citealt{Forster2009}). We find that both clumpy and non-clumpy galaxies have a range of kinematics, but can be viewed as rotationally-dominated systems. Given the upper limits in the dispersion, it is also possible for the non-clumpy sample to have lower dispersion compared to clumpy galaxies.

The sloped line in Figure \ref{fig:kinematics_integrated} is derived from the Toomre's \textit{Q} parameter \citep{Toomre1964}. Instability within a gas disk is predicted when \textit{Q} falls before unity. \textit{Q} is typically expressed as,
\begin{equation} \label{eqn:toomre}
    Q = \frac{\sigma \kappa}{\pi G \Sigma_\mathrm{gas}}, 
\end{equation}

\noindent where $\sigma$ is the local dispersion, $\kappa$ is the epicyclic frequency, and $\Sigma_\mathrm{gas}$ is the gas density. $\kappa$ is related to the angular velocity $\Omega$ such that $\kappa = a \Omega$, and $a$ ranges from 1 for a Keplerian orbit to $\sqrt{3}$ for a uniform disk \citep{Dekel2009a}. As shown by \cite{Genzel2011}, the Toomre parameter can then be reduced to, 

\begin{equation} \label{eqn:toomre_reduced}
    \begin{split}
        Q &= \frac{\sigma \kappa}{\pi G \Sigma_\mathrm{gas}} =  \frac{\sigma (a v_c/R_\mathrm{disk})}{\pi G \Sigma_\mathrm{gas}} \\
        &= a \Big(\frac{\sigma}{v_c}\Big) \Big(\frac{v_c^2 R_\mathrm{disk}/G}{\pi R_\mathrm{disk}^2 \Sigma_\mathrm{gas}} \Big)
        = a \Big(\frac{\sigma}{v_c}\Big) \Big(\frac{M_\mathrm{tot}}{M_\mathrm{gas}}\Big) \\
        Q &= a f_{\mathrm{gas}}^{-1} \frac{\sigma}{v_c}, 
    \end{split}
\end{equation}

\noindent and $f_\mathrm{gas}$ is the gas fraction. Assuming a flat rotation with $a = \sqrt{2}$, the following equation is obtained

\begin{equation}\label{eqn:stable_disk}
    \frac{\sigma}{v_c} = Q \frac{f_{\mathrm{gas}}}{\sqrt{2}}. 
\end{equation}
\noindent Setting $Q=1$ for the instabilitiy criterion gives the sloped relation in Figure \ref{fig:kinematics_integrated}. Lower values for $Q$ indicate unstable disks (i.e., the shaded region). Taking Figure \ref{fig:kinematics_integrated} at face value, it appears that non-clumpy galaxies might be more kinematically stable than clumpy galaxies, however, given that many non-clumpy galaxies have only upper limits in the dispersion we do not find a strong preference for clumpy galaxies to have lower Toomre values compared to non-clumpy galaxies. 

It is not surprising that most galaxies are found to be marginally stable with $Q \sim 1$, if we consider the result within the context of a self-regulating disk. 
If the gas is accreted smoothly such as there is no injection of turbulence, $Q<1$ is more readily achieved in the outer region of the disk given the increase in gas density and flatter rotation curve. This naturally leads to the fragmentation and formation of clumps. Several processes can then regulate \textit{Q} by driving up turbulence. Clump-clump interaction can lead to the transport of mass centrally, and the gain in gravitational energy during this process can increase gas turbulence. Similarly, clumpy accretion of gas is also suggested to drive up turbulence. 

It is important to address several caveats in this analysis. We note that the use of $Q$ here only accounts for the gaseous component of the disk. An effective $Q$ that takes into account both the gaseous and stellar component may better predict the stability of the disk. For example, \cite{Puech2010} found that while clumpy galaxies at $z\sim0.6$ have $Q_\mathrm{gas}$ that is stable, their $Q_\mathrm{eff}$ is below unity. Likewise, Eqn. \ref{eqn:toomre} describes the local instability within a gas disk. Since we are assuming a gas fraction that is derived from the integrated SFR, using the Kennicutt-Schmidt relation to estimate the gas fraction may be incorrect as clumpy galaxies have localized star formation. Similarly, the kinematics are measured based on the ionized gas, which can be susceptible to processes such as outflows. It may be insightful to do similar analyses in the future at higher resolution with molecular gas observations, as such observations can provide better constraints.

\section{Discussion} \label{sec:discussion}
\subsection{Gas inflow driving the MZR offset}

The simplest explanation for the metallicity offset between galaxies of similar masses and SFRs is with the accretion of metal-poor gas; either induced by mergers, or cosmological gas accretion. Indeed, simulations have shown that the scatter in the mass-metallicity relation can be reproduced as the variations in SFR and metallicity \citep{Torrey2018}. The fluctuation of the gas supply can play a role in these variations, where lower gas inflow leads to the increase of metallicity \citep{Lucia2020}. Similarly, \cite{Loon2021} found that gas inflow influences the scatter of the mass-metallicity the most compared to other parameters such as outflows and star formation rate in the \texttt{EAGLE} simulation. 

Observationally, lower metallicities in high-redshift ($z\approx3$) galaxies can also be explained through gas accretion. By using a chemical evolution model from \cite{Erb2008}, \cite{Cresci2010} found that the relation between their observed metallicities and the inferred gas fraction is better explained as the accretion of in-falling pristine gas as opposed to outflows due to AGN, or stellar winds. In this model, both the accretion and outflow rate is assumed to be a constant inflow and outflow fraction compared to the SFR. In the former case, this can be viewed as a continuous process or average of discrete accretion event. Similarly, using the MUSE data from the MEGAFLOW survey, \cite{Langan2023} found that galaxies with inflows are typically found at the lower end of the mass-metallicity relation. 


\begin{figure}
    \centering
    \includegraphics[width=\columnwidth]{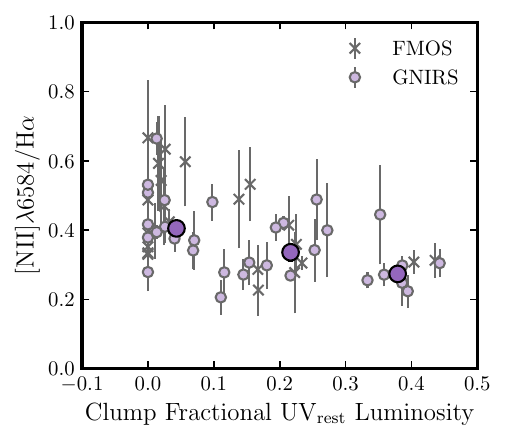}
    \caption{The relation between the {\nii}/{\halpha} line ratio and the fractional UV$_\mathrm{rest}$ contribution for the GNIRS and FMOS samples. The fractional UV$_\mathrm{rest}$ contribution is defined as the ratio of the total rest-frame UV luminosity of all clumps in a galaxy to the total rest-frame UV luminosity of the galaxy. The darker markers show the mean the mean fractional UV$_\mathrm{rest}$ contribution and the weighted mean N2 line ratio for the GNIR sample, using discrete bins along the fractional UV$_\mathrm{rest}$ contribution at between 0-0.15, 0.15-0.3, and 0.3-0.5. Each bin, starting from the lowest bin, has 14, 6, and 6 galaxies. If the fractional UV$_\mathrm{rest}$ contribution is taken as the size/number of clumps in galaxies, the result suggests that clumpier galaxies are more likely to have lower metallicities. }
    \label{fig:niiha_vs_ufrac}
\end{figure}

\subsection{Clump formation due to gas accretion}

Chemical inhomogeneities in the gas disk have been observed both locally and at high-redshift, with localized, kilo-parsec starbursts reportedly having lower metallicities compared to the surrounding regions (e.g., \citealt{Sanchez2015}). In the context of cosmological gas accretion, lower metallicities in such clumpy galaxies can be explained through the accretion of pristine gas. Such metal-poor gas accretion can either triggers a starburst as the gas compresses through its approach to the disk, or results in a build up of the gas disk, which may subsequently form star-forming clumps due to internal instabilities. Using a zoom-in AMR cosmological simulation, \cite{Ceverino2016} found that the accretion of gas leads to the formation of  star-forming clumps in galaxies with $M_* < 10^9 ~M_\odot$. In particular, in all the cases where clumps have a metallicity drop compared to the surrounding interstellar medium, the gas mass of the host galaxies steeply increases, and can double prior to clump formation. Most of their observed clumps have a metallicity drop of around 0.3 dex, and are typically dispersed within a few dynamical times.

\subsection{Clump formation due to mergers}

Wet mergers have also been shown to lead to a dilution of metal contents in galaxies, with major mergers resulting in a larger metallicity drop in comparison to minor mergers \citep{Bustamante2018}. The study found that the metallicity dilution can range from approximately 0.05, 0.1, 0.15 dex for minor (mass ratio $< 1/10$), intermediate ($1/10 <$ mass ratio $< 1/3$), and major (mass ratio $> 1/3$) mergers, respectively. Our observed metallicity offset of $0.06-0.07$ dex for both the GNIRS and FMOS sample suggests that both minor and immediate mergers could be the main driver for clump formation. This is broadly consistent with other studies that suggested that clump formation could be formed to minor mergers at low-redshift. In particular, the major merger rate is relatively low at $z\sim1$ and is therefore not expected to be a major contributor of clump formation, while minor mergers are still relatively common at $z\sim1$ (e.g., \citealt{Guo2015, Shibuya2016}). 
It should also be noted that the galaxy major merger rate is expected to be higher toward high-$z$ \citep{Romano2021, Shibuya2022, Duan2024}. Clumps observed at $z>4$ (e.g., \citealt{Shibuya2016, Ribeiro2017, Hainline2024}) could therefore be formed due to major mergers. In fact, \cite{Nakazato2024} found that most of their detected clumps at the redshift range of 5.5 and 9.5 were induced by mergers using the FirstLight simulation. We note that most of our galaxies are rotationally-dominated, which may imply that the observed metallicity offset is not driven by major mergers. However, such kinematics signatures may not be a good indicator for major mergers as the galaxies may also be in their late merger stage.

\begin{figure}
    \centering
    \includegraphics[width=\columnwidth]{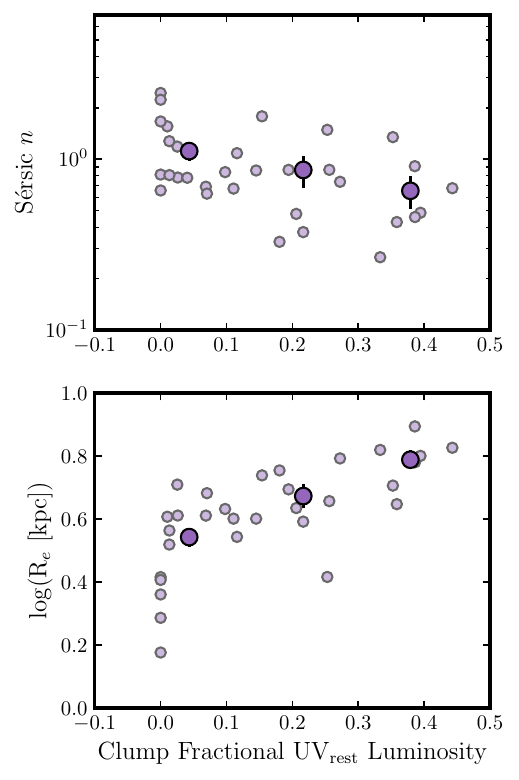}
    \caption{Similar to Fig. \ref{fig:niiha_vs_ufrac}, but is now showing the comparison between the S\'ersic index (top panel) and R$_e$ (bottom panel) to the fractional UV$_\mathrm{rest}$ contribution of clumps. The darker markers now denote the mean, and the error here is taken as the standard error.}
    \label{fig:n_reff_vs_ufrac}
\end{figure}

\subsection{What are the properties of the host SFGs?}

Within the picture of gas inflow, the accreting gas would lead to the dilution of metallicity, and a burst of star formation. This relation between UV-bright, star-forming clumps and lower metallicities is also observed in our results. Figure \ref{fig:niiha_vs_ufrac} show the relation between the mean {\nii}/{\halpha} line ratio as a function of the fractional rest-frame UV contribution from all clumps in a galaxy compared to the integrated value of its host galaxy for both the GNIRS and FMOS samples. The fractional UV contribution can be taken as the size and/or number of the clumps, where higher values can indicate that a galaxy that hosts more and/or massive clumps. We find a slight correlation between the {\nii}/{\halpha} line ratio and fractional rest-frame UV contribution. A Spearman correlation test yields a coefficient of -0.47 with a p-value of approximately 0.006. The result indicates that galaxies with a higher fractional UV contribution from clumps tend to have lower metallicities.

In addition, if gas accretion helps build up the disk of the galaxies, leading to violent disk instability, we would expect the galaxies to have a disk-like morphology. The relation between clumpy and disk-like morphologies is reported by \cite{Shibuya2016}, who observed that the clumpy fraction is higher for galaxies with lower S\'ersic \textit{n} index (i.e., $n\sim1$ indicates a disk-like component). \cite{Martin2023} reported no significant differences between the S\'ersic index between clumpy and non-clumpy galaxies. We perform a similar analysis on our sample. Figure \ref{fig:n_reff_vs_ufrac} shows the relation between the S\'ersic index and the effective radius to the fractional UV contribution. Our analysis indicates that most of our galaxies exhibit disk-like structures. Additionally, we identify a slight correlation between the fractional $UV_\mathrm{rest}$ contribution of clumps and the S\'ersic index of the galaxies. Utilizing the Spearman correlation coefficient test, we obtain a coefficient of -0.44, with a corresponding p-value of 0.01. On the other hand, the fractional UV$_\mathrm{rest}$ contribution of clumps is correlated with the effective radius of the galaxies, with a Spearman coefficient of 0.78 (p-value $< 0.001$), suggesting that clumpier galaxies are larger in size. This can be explained within the framework of violent disk instability (e.g., see \citealt{Dekel2009a}). Firstly, gas accretion is anticipated to accumulate in the outer regions of the disk, leading to increased gas density in these areas. With smooth accretion, gas turbulence would remain unaffected. Finally, assuming a flat rotation curve, the angular velocity is expected to fall off as $\Omega \sim r^{-1}$. Together, these factors naturally lead to a decreasing Toomre parameter, $Q \sim \sigma \Omega / \Sigma$. We note that the relation between the size and fractional UV$_\mathrm{rest}$ contribution could also be an observational artifact where it is easier to identify clumps that are further extended from the central regions of the galaxies.

\section{Conclusion}
Star-forming clumps in high-redshift SFGs are suggested to form as the result of violent disk instabilities within the gas-rich disk. Such disks are believed to be sustained by the continuous accretion of pristine gas from cosmological stream. If pristine gas accretion leads to clump formation due to VDI, we expect to find lower metallicities in clumpy galaxies. 

In this work, we obtain the {\nii}/{\halpha} line ratio for a sample of 32 SFGs at $z\approx0.7$ and 23 SFGs at $z\approx1.5$ in the COSMOS field in order to validate the above scenario by testing whether clumpy galaxies have lower metallicities compared to non-clumpy galaxies. The sample of galaxies consists of both clumpy and non-clumpy galaxies that are selected to have similar stellar masses and star formation rates. The classification of clumpy and non-clumpy is made based on whether the galaxy hosts clumps that account for more than $8\%$ of the total rest-frame UV luminosity of the galaxy. We infer the gas-phase metallicity based on the line ratio of {\nii} and {\halpha}. The main takeaways of the paper are as follows.

\begin{itemize}
    \item We measure the mass-metallicity relation and find that it is in agreement with other studies of the relation at similar redshift ranges. Our weighted metallicity average is $8.67\pm0.01$ dex.
    \item Clumpy galaxies have lower metallicities compared to non-clumpy galaxies, with a difference in the metallicity of $0.07\pm0.02$ dex.
    \item Furthermore, we find a similar metallicity offset at a higher redshift by performing similar analyses using data from the FMOS-COSMOS survey. FMOS-COSMOS contains fluxes for both {\halpha} and {\nii} for galaxies in the COSMOS field at $z\approx1.5$. We apply the same morphological classification to these galaxies, and observe a metallicity offset of $0.06\pm0.02$ dex. 
    \item Both clumpy and non-clumpy galaxies exhibit a wide range of disk kinematics, with $\sigma_0/v_c$ ranging between 0.02 and 0.4. This indicates that the majority of our SFGs are rotationally-dominated.
    \item We do not find a significant difference in the kinematics of clumpy and non-clumpy galaxies. 
    \item The {\nii}/{\halpha} ratio is negatively correlated with the fractional $\mathrm{UV_{rest}}$ luminosity contribution from clumps. Interpreting the fractional $\mathrm{UV_{rest}}$ luminosity as the number and/or size of clumps, this result suggests that \enquote{clumpier} galaxies tend to have lower metallicities.
\end{itemize}

\noindent Our results indicate a metallicity offset between clumpy and non-clumpy galaxies, as well as a clear relation between metallicity and the number of clumps. These results suggest that clump formation is driven by the inflow of metal-poor gas, and that the processes that form them may not require significant kinematics instability in the gas disk. However, given the observational uncertainties, performing similar analyses with direct molecular gas observations would be valuable to better constraints for these parameters.

\section{Acknowledgments}
The authors would like to thank the referee for the constructive feedback. GW gratefully acknowledges support from the National Science Foundation through grant AST-2205189 and from HST program number GO-16300. This paper is based on observations obtained at the international Gemini Observatory, a program of NSF’s NOIRLab, which is managed by the Association of Universities for Research in Astronomy (AURA) under a cooperative agreement with the National Science Foundation on behalf of the Gemini Observatory partnership: the National Science Foundation (United States), National Research Council (Canada), Agencia Nacional de Investigaci\'{o}n y Desarrollo (Chile), Ministerio de Ciencia, Tecnolog\'{i}a e Innovaci\'{o}n (Argentina), Minist\'{e}rio da Ci\^{e}ncia, Tecnologia, Inova\c{c}\~{o}es e Comunica\c{c}\~{o}es (Brazil), and Korea Astronomy and Space Science Institute (Republic of Korea).


%

\vspace{5mm}
\facility{Gemini-North (GNIRS)}


\software{\texttt{astropy} \citep{astropy2013, astropy2018, astropy2022}, \texttt{emcee} \citep{Foreman2013}, \texttt{lmfit} \citep{Newville2016}, \texttt{specutils} \citep{Earl2022}}, \texttt{statmorph} \citep{Rodriguez2019}

\begin{longrotatetable}
\setlength\extrarowheight{-0.2pt}
\begin{deluxetable*}{ccccccccccc}
\tablecaption{Derived values for the GNIRS observations. We only list the values for galaxies with detected {\halpha} and {\nii} lines. The {\nii}/{\halpha} line ratio is used to obtain the gas-phase metallicity using the calibration from \cite{Pettini2004}. }
\tablehead{
    \colhead{COSMOS ID} & 
    \colhead{RA} &
    \colhead{Dec} &
    \dcolhead{z_\mathrm{spec}} &
    \dcolhead{\log(\mathrm{M_*})} &
    \dcolhead{\log(\mathrm{SFR_{UV+IR}})} & 
    \colhead{UV$_\mathrm{clumpy}$} & 
    \colhead{\nii/\halpha} & 
    \dcolhead{12 + \log(\mathrm{O/H})} & 
    \dcolhead{v_c} &
    \dcolhead{\sigma_0} \\
 \colhead{} &
 \colhead{[deg]} &
 \colhead{[deg]} &
 \colhead{} &
 \colhead{$[\log(\mathrm{M_\odot})]$} &
 \colhead{$[\log(\mathrm{M_\odot}~yr^{-1})]$} &
 \colhead{} &
 \colhead{} &
 \colhead{} &
 \dcolhead{[\mathrm{km ~s^{-1}}]} &
 \dcolhead{[\mathrm{km ~s^{-1}}]} }
\startdata
2289 & 150.48253 & 1.7125193 & 0.712 & 10.44 & 0.99 & True & 0.26 $\pm$ 0.02 & 8.56 $\pm$ 0.02 & 172.3 $\pm$ 14.8 & 13.7 $\pm$ 6.1 \\
40738 & 150.24486 & 1.9678429 & 0.936 & 10.44 & 1.45 & True & 0.21 $\pm$ 0.05 & 8.51 $\pm$ 0.06 &  \nodata  & 35.8 $\pm$ 23.7 \\
62945 & 149.72485 & 1.8760713 & 0.694 & 10.44 & 1.09 & True & 0.28 $\pm$ 0.07 & 8.58 $\pm$ 0.06 &  \nodata  & \nodata  \\
64167 & 149.66539 & 1.8908899 & 0.749 & 10.48 & 0.99 & False & 0.28 $\pm$ 0.06 & 8.58 $\pm$ 0.05 &  \nodata  & \nodata  \\
73816 & 150.5074 & 2.0096676 & 0.702 & 10.59 & 1.45 & False & 0.38 $\pm$ 0.04 & 8.66 $\pm$ 0.02 & 212.5 $\pm$ 22.2 & 4.1 $\pm$ 7.5 \\
85093 & 150.67809 & 2.1846418 & 0.795 & 10.52 & 1.08 & True & 0.44 $\pm$ 0.14 & 8.70 $\pm$ 0.08 &  \nodata  & \nodata  \\
91389 & 150.72302 & 2.2811966 & 0.957 & 10.41 & 1.18 & True & 0.25 $\pm$ 0.07 & 8.55 $\pm$ 0.07 &  \nodata  & 26.3 $\pm$ 17.1 \\
91840 & 150.68439 & 2.2863963 & 0.607 & 10.61 & 1.08 & True & 0.49 $\pm$ 0.12 & 8.72 $\pm$ 0.06 &  \nodata  & 21.5 $\pm$ 13.9 \\
94152 & 150.4547 & 2.32392 & 0.662 & 10.58 & 0.93 & False & 0.37 $\pm$ 0.08 & 8.65 $\pm$ 0.06 & 186.7 $\pm$ 46.3 & 37.3 $\pm$ 20.4 \\
111826 & 150.16794 & 2.099118 & 0.961 & 10.65 & 1.41 & False & 0.40 $\pm$ 0.08 & 8.67 $\pm$ 0.05 &  \nodata  & 23.3 $\pm$ 15.2 \\
116791 & 150.0715 & 2.1623764 & 0.662 & 10.57 & 1.53 & False & 0.42 $\pm$ 0.01 & 8.68 $\pm$ 0.01 & 108.2 $\pm$ 13.7 & 23.3 $\pm$ 7.0 \\
118818 & 150.15173 & 2.1886752 & 0.927 & 10.56 & 1.70 & True & 0.27 $\pm$ 0.04 & 8.58 $\pm$ 0.04 & 151.5 $\pm$ 17.8 & 54.6 $\pm$ 7.5 \\
118902 & 150.15822 & 2.1901271 & 0.925 & 10.51 & 1.57 & False & 0.34 $\pm$ 0.05 & 8.63 $\pm$ 0.04 & 179.7 $\pm$ 37.6 & 17.4 $\pm$ 12.4 \\
120517 & 150.18661 & 2.2085881 & 0.892 & 10.47 & 1.52 & False & 0.41 $\pm$ 0.05 & 8.68 $\pm$ 0.03 & 201.8 $\pm$ 30.1 & 79.4 $\pm$ 13.1 \\
121358 & 150.17503 & 2.2170861 & 0.679 & 10.42 & 1.31 & True & 0.27 $\pm$ 0.02 & 8.57 $\pm$ 0.01 & 173.3 $\pm$ 28.7 & 35.7 $\pm$ 6.9 \\
122592 & 150.14954 & 2.2334945 & 0.731 & 10.35 & 1.16 & True & 0.27 $\pm$ 0.03 & 8.58 $\pm$ 0.03 & 68.2 $\pm$ 15.0 & 21.3 $\pm$ 7.7 \\
122650 & 150.07106 & 2.2330399 & 0.603 & 10.53 & 1.28 & False & 0.39 $\pm$ 0.02 & 8.67 $\pm$ 0.01 & 167.4 $\pm$ 20.6 & 14.8 $\pm$ 6.7 \\
123666 & 150.21819 & 2.2458375 & 0.657 & 10.61 & 1.36 & True & 0.42 $\pm$ 0.02 & 8.69 $\pm$ 0.01 & 255.1 $\pm$ 16.3 & 40.0 $\pm$ 10.6 \\
125867 & 150.36453 & 2.2735415 & 0.749 & 10.41 & 1.24 & True & 0.22 $\pm$ 0.05 & 8.53 $\pm$ 0.05 & 127.1 $\pm$ 36.3 & 32.1 $\pm$ 12.4 \\
127326 & 150.17822 & 2.292134 & 0.748 & 10.42 & 1.40 & True & 0.34 $\pm$ 0.09 & 8.63 $\pm$ 0.06 &  \nodata  & 49.9 $\pm$ 20.1 \\
151806 & 149.88297 & 2.1179976 & 0.677 & 10.49 & 0.94 & True & 0.30 $\pm$ 0.04 & 8.61 $\pm$ 0.03 & 149.4 $\pm$ 15.1 & 28.4 $\pm$ 8.6 \\
153139 & 149.73523 & 2.1352146 & 0.698 & 10.43 & 1.18 & True & 0.41 $\pm$ 0.04 & 8.68 $\pm$ 0.02 & 171.7 $\pm$ 17.1 & 33.4 $\pm$ 11.0 \\
159971 & 149.61758 & 2.215559 & 0.675 & 10.54 & 1.12 & False & 0.49 $\pm$ 0.06 & 8.72 $\pm$ 0.03 & 190.9 $\pm$ 19.4 & 16.2 $\pm$ 11.8 \\
161993 & 149.8849 & 2.2384937 & 0.703 & 10.55 & 1.40 & False & 0.66 $\pm$ 0.05 & 8.80 $\pm$ 0.02 &  \nodata  & \nodata  \\
162807 & 149.75438 & 2.2461481 & 0.730 & 10.59 & 1.21 & True & 0.40 $\pm$ 0.14 & 8.67 $\pm$ 0.08 &  \nodata  & \nodata  \\
185948 & 150.51099 & 2.5288169 & 0.852 & 10.56 & 1.19 & False & 0.38 $\pm$ 0.15 & 8.66 $\pm$ 0.09 &  \nodata  & \nodata  \\
191595 & 150.53107 & 2.6109066 & 0.815 & 10.53 & 1.40 & False & 0.51 $\pm$ 0.09 & 8.73 $\pm$ 0.04 &  \nodata  & \nodata  \\
192043 & 150.65984 & 2.6178505 & 0.891 & 10.5 & 1.23 & True & 0.31 $\pm$ 0.07 & 8.61 $\pm$ 0.05 &  \nodata  & 35.6 $\pm$ 23.1 \\
207078 & 149.98328 & 2.4797268 & 0.704 & 10.5 & 1.40 & True & 0.30 $\pm$ 0.03 & 8.60 $\pm$ 0.02 & 75.9 $\pm$ 26.2 & 20.0 $\pm$ 9.2 \\
207362 & 150.17223 & 2.4845951 & 0.800 & 10.48 & 1.24 & False & 0.53 $\pm$ 0.09 & 8.74 $\pm$ 0.04 &  \nodata  & \nodata  \\
239578 & 149.84 & 2.5136211 & 0.679 & 10.6 & 1.43 & False & 0.48 $\pm$ 0.05 & 8.72 $\pm$ 0.03 & 220.0 $\pm$ 26.6 & 15.6 $\pm$ 11.2 \\
239638 & 149.72881 & 2.515331 & 0.696 & 10.43 & 1.03 & True & 0.30 $\pm$ 0.07 & 8.60 $\pm$ 0.06 & 145.6 $\pm$ 23.1 & 38.8 $\pm$ 14.7 \\
\enddata
\label{tab:gnirs}
\end{deluxetable*}
\end{longrotatetable}


\bibliography{reference}{}
\bibliographystyle{aasjournal}



\end{document}